**Radu RUSU, PhD Candidate**
E-mail: raduav.rusu@ulbsibiu.ro
Lucian Blaga University of Sibiu, Romania

**Professor Camelia OPREAN-STAN, PhD (corresponding author)**
E-mail: camelia.oprean@ulbsibiu.ro
Lucian Blaga University of Sibiu, Romania

# THE IMPACT OF DIGITALISATION AND SUSTAINABILITY ON INCLUSIVENESS: INCLUSIVE GROWTH DETERMINANTS

***Abstract.*** *Inclusiveness and economic development have been slowed by the pandemics and military conflicts. This study investigates the main determinants of inclusiveness at the European level. A multi-method approach is used, with Principal Component Analysis (PCA) applied to create the Inclusiveness Index and Generalised Method of Moments (GMM) analysis used to investigate the determinants of inclusiveness. The data comprises a range of 22 years, from 2000 to 2021, for 32 European countries. The determinants of inclusiveness and their effects were identified. First, economic growth, industrial upgrading, electricity consumption, digitalisation, and the quantitative aspect of governance, all have a positive impact on inclusive growth in Europe. Second, the level of $CO_2$ emissions and inflation have a negative impact on inclusiveness. Tomorrow's inclusive and sustainable growth must include investments in renewable energy, digital infrastructure, inequality policies, sustainable governance, human capital, and inflation management. These findings can help decision makers design inclusive growth policies.*

## 1. Introduction

In recent years, there has been a growing interest in inclusiveness as a multidimensional concept that can relate to equity, empowerment, opportunities, participation, satisfaction, or a combination of these elements (Hay et al., 2022). Inclusiveness is the concept of offering equal economic and social opportunities to all segments of society, regardless of wealth or social status. The notion of inclusive growth has gained importance in development circles, but there is still no consensus on what it truly means, and the dispute over its definition continues.

There are also challenges in measuring inclusiveness, and a comprehensive framework that captures its multidimensional nature is required. Within economically developed countries, the determinants of inclusive growth have not yet been adequately recognised and analysed.

The goal of this research is to fill these gaps by investigating the fundamental determinants of inclusiveness at the European level and assessing the importance of the main determinants in achieving inclusiveness for the countries in the sample. In addition, this paper proposes the development of an inclusiveness index, where the choice of component indicators was made with respect to the economic, environment and social interconnectivity. This inclusiveness index can be used as a tool for policymakers to identify areas of improvement in their countries and to develop policies that promote inclusive growth.

Also, given its multiple links with green economic growth and general sustainability, digitalisation is an important factor of inclusiveness (Wu et al., 2021; Stan et al., 2020). As a result of the proliferation of telecommunication infrastructure and the internet, issues such as time and distance can be better managed, allowing for the completion of more complex tasks.

Another phenomenon to consider is inflation, which has a significant and negative impact on economic growth, as indicated by several studies (e.g., Ayyoub et al., 2021). However, financial inclusion can help decrease the inflation rate in developing countries, thereby promoting inclusive growth. The impact of inflation on growth is also found to vary depending on the strength of institutions, with negative effects observed in countries with stronger institutions and positive effects in countries with weaker institutions.

The present study uses a multi-method approach combining principal component analysis (PCA), for the creation of the Inclusiveness Index based on dimension reduction, and Generalised Method of Moments (GMM) analysis to study the determinants of inclusiveness. The current paper employs 32 European countries during a 22-year period, as well as a set of metrics appropriate for the objective (Ofori et al., 2022; Wang et al., 2022). The findings successfully identified the inclusiveness determinants and the nature of their influence. Investments in renewable energy technologies, digital infrastructure, inequality policies, and sustainable governance will be characteristics of tomorrow's inclusive and sustainable growth.

This article provides valuable insights for future research on inclusiveness and inclusive growth at the European level. The findings should contribute significantly to understanding how the COVID-19 pandemic and the Russo-Ukrainian war are having a negative impact on European welfare. This study aims to add to the existing knowledge on inclusiveness and to propose improved solutions for inclusive growth.

The remaining part of the paper proceeds as follows: Section 2 introduces the literature review related to the environmental side of inclusiveness, but also to the social dimension and economic determinants of inclusive growth. Section 3 is concerned with the methodology used for this study for generating the Inclusiveness Index and for analysing the principal determinants of inclusiveness. The fourth section presents the findings of the research. The conclusions, limitations, and directions for future studies are summarised in Section 5.

## 2. Literature Review

### 2.1. *Environmental side of Inclusiveness*

Practically, the current research seeks to identify and analyse the primary factors that determine inclusiveness and the environment-related features of inclusive growth (Ding et al., 2023; Wang et al., 2022). The environmental side of inclusiveness comprises elements such as: the extent by which renewable energy is consumed (Ofori et al., 2022), innovation related to new technologies that address energy generated from renewable sources (Razzaq et al., 2023), energy consumption together with $CO_2$ emissions, the structure of the resources used in generating primary energy (Koengkan et al., 2021) and the negative effects on the population's welfare represented by fine particles (PM2.5) that are a result of combustion activities (Ofori et al., 2022; Wang et al., 2022). PM2.5 are responsible for increasing the extent to which mortalities occur and for increasing the welfare costs implied by the prevention of such deaths caused by the exposure to the ambient particles (Ofori et al., 2022).

With regard to achieving inclusive growth, the following strategies are worth taking into account. First of all, the extent to which effective governance is specific to a country has been shown to positively influence inclusiveness (Ofori et al., 2022). In addition to this, the six dimensions of governance, as stated by World Governance Indicators, have meaningful and accentuated impacts in determining inclusiveness given the fact that they appear to be diminishing the income discrepancies between the poor and the rich (Ofori et al., 2022). On the other hand, governance has been shown to negatively influence the degree of greenhouse gas emissions (Li et al., 2023; Lee et al., 2022; Ofori et al., 2022). Apart from its individual influences, governance positively influences inclusiveness when paired with energy efficiency (Ofori et al., 2022; Wang et al., 2022).

To this end, the present paper sets out to test the following hypotheses:

*H1*: The quantity of electricity consumed per capita exerts a positive influence on the degree of inclusiveness.

*H2*: The degree of $CO_2$ emissions has a negative impact on inclusiveness.

*H3*: Inclusiveness is positively influenced by both the quantitative and the qualitative sides of the governance dimension.

Radu Rusu, Camelia Oprean-Stan### 2.2. Social Dimension Characteristic to Inclusiveness

Digitalisation has been proved to be directly associated to inclusive growth, as seen by the number of internet users (Shodiev et al., 2021). In addition, the amount to which private companies have access to digital funds positively influences their judgments towards environmental protection (Razzaq et al., 2023). Human capital, as part of the digitalisation index (Wang et al., 2022), has been shown to be important determinant of green economic growth. Consequently, a quantitative measure of human capital in regressing inclusiveness demonstrates its validity based on past research.

Concerning the existing literature, the following hypotheses have been set for testing:

*H4*: Human capital is positively influencing the degree of inclusiveness.

*H5*: The level of digitalisation accounts for a positive impact toward inclusiveness.

### 2.3 Economic Determinants of Inclusiveness

The socio-economic context reflects the welfare of the population. As a consequence, inflation as a macroeconomic indicator is a negative determinant of inclusive growth (Ofori et al., 2022). The effect of inflation that results in a decrease in purchasing power has a greater impact on the poor than on the wealthy. This causes certain social categories to be unable to afford essential goods and services. This reality is specific to developing countries, where economic stability is not substantially established. Hence, it is innovative to examine the impact of inflation on inclusiveness at the European level in the present study.

Inclusiveness has been shown to be positively influenced by the extent in which a country benefits from foreign financial aid, and foreign aid has been shown to positively influence the extent by which gas emissions occur (Ofori et al., 2022). Given that the cross-sections used in the present study imply developed countries, the degree of economic development shall be reflected through the means of gross domestic product per capita and industrial upgrading. Economic growth has been shown to exert a positive impact on the degree of inclusiveness, by diminishing the income discrepancies between the population of developing countries. In addition to this, an increase in the degree of economic development is shown to exert a negative effect on the extent of greenhouse gas emissions. Furthermore, economic growth means more resources can be invested in research and development activities so that innovation can take place both in the field of renewable technologies and that of digitalisation (Razzaq et al., 2023).

Therefore, the last hypotheses to be tested in the current paper are built:

*H6*: Economic development, reflected by gross domestic product per capita and industrial upgrading, has a positive impact on inclusiveness.

*H7*: Inflation exerts a negative influence on inclusiveness.

212

The Impact of Digitalisation and Sustainability on Inclusiveness:
Inclusive Growth Determinants

## 3. Methodology

### *3.1 Generating the Inclusiveness Index*

The present paper uses macroeconomic data for a number of 32 European countries for 22-year range that spans from 2000 to 2021. The European context is appropriate for investigation because the current study will provide insights into how various macro level phenomena affected the extent of inclusiveness within a region characterised by advanced economic development. In addition, not enough attention has been paid in the empirical literature to this region. The large period, namely from 2000 to 2021, makes the context more comprehensive, as it offers a broader perspective into the dynamic of inclusiveness and inclusive growth at the European level. The dependent variable, inclusiveness, is at the core of the following analyses and is not available in online databases, so it must be generated. Dimension reduction using principal component analysis is the method used to accomplish this and follows the techniques applied in previous research by Ofori et al. (2022) and Tchamyou et al. (2019).

As for generating the inclusiveness index, the choice of component indicators was made with respect to the economic, environment, and social interconnectivity. As a consequence, 15 emblematic covariates (Table 1) that are specific to the three-dimensional interconnectivity have been identified and included in the principal component analysis. The social aspect of the three-dimensional interconnectivity is reflected by elements such as education, unemployment, and the population access to safely managed services (Ofori et al., 2022). The environment related aspects are reflected by the structure of the land, agricultural and forests, as well as by the degree of exposure to PM2.5 microparticles. Their importance resides in the fact that the exposure to such particles results in additional welfare costs meant to mitigate their negative effects, but also in premature deaths that have the particles as the primary causes. The economic dimension is reflected in the advancements that have been made so far and are reflected by the overall population's access and use of energy and its active participation in the work force.

In order to obtain the inclusiveness index, Principal Component Analysis (PCA) proved to be the answer to such a task, prior to the works of Amavilah et al, (2017) and Tchamyou et al. (2019). The advantage of PCA is that it reduces the number of variables initially used by taking into account a small number of representative variables, i.e., reducing large data volume to a structure that retains as much common variability as possible. The three-dimensionality of the chosen set of indicators will be kept while their numbers will be reduced, thus obtaining the principal components. The statistical significance of the PCA is stated in Table 2. The value obtained for the KMO measure of sampling adequacy, which is greater than 0.50, and the high value of the Bartlett Chi-square, together with the statistical significance of the p-value, are a statement of the interconnection between the 15 covariates used for generating inclusiveness.

Radu Rusu, Camelia Oprean-Stan**Table 1. Variables used in determining the inclusiveness index**

| Indicator | Symbol | Definition | Source |
|---|---|---|---|
| Agricultural area | AGR | % of the total land area represented by agricultural land | WDI |
| Child mortality rate | CMR | The number of child deaths per 1000 children | OWD |
| Clean fuels | CFT | Population's access to clean fuels and technologies for cooking | WDI |
| Electricity net consumption | ENC | The quantity of electricity consumed | WDI |
| Exposure to PM25 | PM25 | The mean population's exposure to microparticles | OWD |
| Forest area | FA | % of the total land area represented by forests | WDI |
| Labor participation female | LPF | Percentage of the female population active in the labour force | WDI |
| Labor participation male | LPM | Percentage of the male population active in the labour force | WDI |
| Life expectancy at birth | LFE | The child's life expectancy at birth, if all elements are unchanged during the course of the life | HDI |
| Parliament seats female | PSF | The number of seats held in the parliament by female | WDI |
| Population with at least secondary education, female | SEF | The percentage of the female population that has attained at least secondary education | WDI |
| Population with at least secondary education, male | SEM | The percentage of the male population that has attained at least secondary education | WDI |
| Total dependency ratio | TDR | The ratio of youth and elderly population per 100 active workers of age 15-64 | OWD |
| Unemployment with advanced education | UAE | The percentage of population that is unemployed and has attained advanced levels of education | WDI |
| Unemployment total | UNT | The total ratio of unemployment, regardless of the educational level attained | WDI |

Source: Author's construct based on available data, 2023
*WDI stands for World Data Indicators, OWD stands for Our World in Data, and HDI stand for Human Development Index

Before conducting the PCA, given that the 15 indicators come in various scales, the data have been normalised so that they present a mean of 0 and a standard deviation of 1 (as in Ofori et al., 2022). After conducting the PCA, in light of previous research in the field (Tchamyou et al., 2019), the inclusiveness index was generated through the means of the components that present an eigenvalue greater than 1, as can be seen from Table 3.

**Table 2. KMO and Bartlett's sphericity test**

| KMO and Bartlett's Test | | |
|---|---|---|
| Kaiser-Meyer-Olkin Measure of Sampling Adequacy. | | 0.639 |
| Bartlett's Test of Sphericity | Approx. Chi-Square | 8072.59 |
| | df | 105 |
| | Sig. | 0 |

214

**Table 3. Principal components' eigenvalues**

| | Total Variance Explained | | | | | | | | |
|---|---|---|---|---|---|---|---|---|---|
| Component | Initial Eigenvalues | | | Extraction Sums of Squared Loadings | | | Rotation Sums of Squared Loadings | | |
| | Total | % of Var. | Cumul% | Total | % of Var. | Cumul % | Total | % of Var. | Cumul % |
| **1** | **4.854** | **32.358** | **32.358** | **4.854** | **32.358** | **32.358** | **3.463** | **23.083** | **23.083** |
| **2** | **2.254** | **15.025** | **47.383** | **2.254** | **15.025** | **47.383** | **2.537** | **16.916** | **39.999** |
| **3** | **1.698** | **11.323** | **58.705** | **1.698** | **11.323** | **58.705** | **1.887** | **12.583** | **52.582** |
| **4** | **1.444** | **9.628** | **68.334** | **1.444** | **9.628** | **68.334** | **1.823** | **12.152** | **64.734** |
| **5** | **1.101** | **7.340** | **75.674** | **1.101** | **7.340** | **75.674** | **1.641** | **10.940** | **75.674** |
| 6 | 0.980 | 6.533 | 82.207 | | | | | | |
| 7 | 0.847 | 5.648 | 87.855 | | | | | | |
| 8 | 0.505 | 3.366 | 91.221 | | | | | | |
| 9 | 0.394 | 2.626 | 93.847 | | | | | | |
| 10 | 0.276 | 1.843 | 95.690 | | | | | | |
| 11 | 0.248 | 1.652 | 97.342 | | | | | | |
| 12 | 0.188 | 1.251 | 98.593 | | | | | | |
| 13 | 0.111 | 0.737 | 99.329 | | | | | | |
| 14 | 0.085 | 0.564 | 99.894 | | | | | | |
| 15 | 0.016 | 0.106 | 100.000 | | | | | | |

Extraction Method: Principal Component Analysis.
Source: Author's construct based on available data, 2023

### *3.2 Establishing the Control Variables and the Principal Determinants*

The main variables are presented in Table 4. The control variables employed in the following analyses are represented by: gross domestic product per capita, which is used in order to control for economic development and is relevant in quantifying the extent by which a country may afford to conduct investments related to the socio-environment dimensions specific to inclusiveness. Such investments refer to the welfare of the population and can address social protection policies, health and education expenditure, employment strategies, and policies. As far as the environment dimension is concerned, with regard to economic development, increases in gross domestic product may imply increases in the share of funds allocated toward research and development in the field of energy-related technologies, carbon sequestration technologies, and so forth. Following economic development, industrial upgrading is the second control variable used and is relevant in outlining the transitions underwent from agriculture to industry and then from services to research-oriented sectors. Because there are no data in online databases that reflect the value added by the research sector to the gross domestic product, the current empirical study will only reflect industrial upgrading through the value added by the services sector. The services sector represents a picture of a transition from labour-intensive work to knowledge-based activities.

Radu Rusu, Camelia Oprean-Stan**Table 4. Description of the main variables and their data sources**

| Variable | Symbol | Definition | Source |
|---|---|---|---|
| $CO_2$ emissions billion tone/capita | $CO_2$ | Annual billion tonnes of $CO_2$ emissions per capita | OWD |
| Economic growth | GDP | The gross domestic product per capita expressed in 2015 USD | WDI |
| Electricity consumption mWh/capita | EC | mWh of electricity consumed per capita | EIA |
| Governance effectiveness | GOV | World Governance Indicators - Control of corruption; Rule of law; Government effectiveness; Regulatory quality; Political stability; Voice & accountability | WGI |
| Government consumption expenditure | GCE | General government final consumption expenditure (% of GDP). | WDI |
| Human capital. | HC | The number of years of schooling a child is expected to attain at birth, if all elements maintain during the course of the life | HDI |
| Inclusiveness | INCL | Inclusiveness index generated by the author using PCA | Author |
| Individuals using the Internet (% of population) | DIG | The share of the individuals that used the internet of the total population | WDI |
| Industrial upgrading | UPGR | The share of the services sector's added value as percentage of gross domestic product | WDI |
| Inflation | INFL | The cumulative inflation, where 1999 represents 100%, up to 2021 | WDI |
| Globalisation | GLO | The index of economic globalisation as expressed by KOFGI | KOFGI |
| Population density | PD | The number of inhabitants per 1 square km of land | OWD |
| Primary energy consumption mWh/capita | PEC | mWh of primary energy consumed per capita | EIA |
| Trade | TRD | Trade is the sum of exports and imports of goods and services measured as a share of gross domestic product. | WDI |
| Urban population (% of total population) | URB | The number of the population that lives in urban areas per total population | WDI |

Source: Author's construct based on the available data, 2023
*KOFGI stands for: Konjunkturforschungsstelle Globalisation Index, EIA stands for: U.S. Energy Information Administration, WGI stands for: World Governance Indicators.

The third control variable used is represented by the quantity of electricity consumed per capita. To some extent, this choice implies an intersection between economic, environment and research aspects given that electricity consumption is a result of economic growth (Bayar et al., 2019). Pollution results from activities that imply electricity consumption and research activities are expected to reduce the negative effects of generating electricity, while simultaneously fully maintaining the desired effects. On the other hand, electricity consumption is a primary condition for digitalisation, which in turn depends on the population's skills, abilities and knowledge related to information and communication technologies.

The Impact of Digitalisation and Sustainability on Inclusiveness:
Inclusive Growth Determinants

The expected years of schooling quantitatively reflect human capital, which in turn comprises the fourth control variable used in the current research. Human capital embodies governmental expenses with education and health for a broader span of years and is directly related to the degree of human development within a country.

Based on previous empirical studies, the current paper individually addresses the impact of four principal determinants on inclusiveness. The first one is represented by the individuals using the Internet as a percentage of the total population. Although a quantitative indicator of digitalisation rather than a qualitative one, the extent of Internet use has been widely used in either generating digitalisation indexes, or in regressing inclusive growth (for example, Ghouse et al., 2022). The construction of a more comprehensive digitalisation indicator was not possible due to inexistent data for the selected cross-sections. In addition, the existing indicator Digital Economy and Society Index does not present available data for every year from the period 2000-2021. Nevertheless, the use of Internet has been shown to negatively influence the extent of pollution and the income discrepancies (Dong et al., 2022; Ofori et al., 2022), while positively influencing green and inclusive growth (Ghouse et al., 2022; Wang et al., 2022).

The second principal determinant is represented by inflation. Paired with gross domestic product, the two highlight the real economic growth compared to the nominal economic growth. Inflation appears to negatively influence inclusiveness in developing countries (Ofori et al., 2022), which is why its inclusion in this study at the level of developed countries in Europe is of interest. Given that European policies and economic development project an image of stability, inflation may not have the same negative impact on Europe as it does on developing nations. Inflation is determined in part by political stability and effectiveness; therefore, governance is the third determinant addressed. The choice of indicators for this element comprises the average of the six components of the world governance indicators, these being corruption control, voice and accountability, government effectiveness, regulatory quality, political stability, and the rule of law. The average scores obtained by each of the 32 countries for the entire period represent the qualitative side of governance, whereas the general government final consumption expenditure reflects the quantitative one. Governance effectiveness and consumption have been widely used in analysing inclusive growth, and the findings state a positive influence both when taken separately (Dong et al., 2022, Wang et al., 2022) and when combined with other indicators, such as energy efficiency (Ofori et al., 2022). In order to have a more comprehensive analysis regarding the environment dimension of inclusive growth, the fourth principal determinant refers to the $CO_2$ emissions. In light of the inclusive index's composition in the present study, regressing through the $CO_2$ emissions will determine to what extent pollution affects inclusiveness with regard to the negative effects of the PM25 on the population's welfare (Koengkan et al., 2021). To the authors' knowledge, there is not sufficient research that targets the influences of pollution on inclusiveness, but rather what factors are responsible for the reduction of $CO_2$ emissions (as in Dong et al., 2022; Ozturk et al., 2022).

### *3.3 Model Specification for the Generalised Method of Moments Analysis*

In addition, the specification of the empirical model of analysis comprises the four control variables and the four principal determinants. A number of 15 variables are employed in the Generalised Method of Moments (GMM) analyses. GMM analysis provides a powerful and flexible approach to estimating parameters in a wide range of statistical and econometric models. The advantages of GMM analysis include robustness to outliers and the ability to handle instrumental variables. Therefore, the following equation is set, where inclusiveness is a function comprising three-dimensional elements as well as error terms:

$$incl_{i,t} = f(\delta_c, \alpha_k, \mu_i, \mu_t, \varepsilon_{i,t}) \tag{1}$$

where $\delta_c$ stands for control variable, $\alpha_k$ stands for principal determinant, $\mu_i$ stands for the effect specific to the country $i$, $\mu_t$ represents the effect specific to the year $t$ and lastly $\varepsilon_{i,t}$ comprises the idiosyncratic error term for the observation of country $i$ at time $t$.

$$incl_{i,t} = \beta_0 + \sum_{c=1}^{4} \beta_c \delta_{c_{i,t}} + \sum_{k=1}^{5} \beta_k \alpha_{k_{i,t}} + \mu_i + \mu_t + \varepsilon_{i,t} \tag{2}$$

where $\beta_{0,c,k}$ are the constant, the control variables' coefficients and the principal determinants' coefficients respectively.

$$incl_{i,t} = \beta_0 + incl_{i,t-1} + \beta_1 gdp_{i,t} + \beta_2 upgr_{i,t} + \beta_3 ec_{i,t} + \beta_4 hc_{i,t} + \sum_{k=1}^{5} \beta_k \alpha_{k_{i,t}} + \mu_i + \mu_t + \varepsilon_{i,t} \tag{3}$$

where the variables used are those in Table 4.

Prior to the specification of the model, all indicators' values have been normalised so they present an average of 0 (zero) and a standard deviation of 1. Further on, in order to proceed to the natural logarithmation, the values were transformed so that the minimum was a positive number.

$$x_{i,t} \Rightarrow x_{i,t_{normalized}}, \text{ with } \mu = 0 \text{ and } \sigma = 1 \tag{4}$$

where $x_{i,t}$ are the observations for the variables presented in Table 4.

$$x_{i,t_{normalized}} \Rightarrow \ln(x_{i,t_{normalized}} + 4) \tag{5}$$

where $\mu = \ln(4)$, $\sigma \in [0.18; 0.38]$ and the value of 4 satisfying the condition for positivity for all the 15 variables used.

The Impact of Digitalisation and Sustainability on Inclusiveness:
Inclusive Growth Determinants

### 4. Results

This section contains the main findings concerning the determinants of inclusiveness. Mentioned in the previous parts, based on equation (3), the chosen models for the GMM analyses consist of four control variables: gross domestic product, industrial upgrading, electricity consumption and human capital; four principal determinants addressed either individually or paired together: digitalisation, inflation, governance and government consumption expenditure, $CO_2$ emissions and also the first lag of the dependent variable.

Based on the processed data presented in Graph 1, negative situations regarding the extent of inclusiveness can be observed in the following countries: Bulgaria, Cyprus, Czechia, Latvia, Lithuania, Moldova, Poland, Romania, Russian Federation, Serbia, Slovak Republic, Slovenia, and Turkey. To some extent, most of the above countries have similarities regarding the influence of the past political regimes on their overall socio-economic status. In contrast, Denmark, Finland, France, Iceland, Norway, Portugal, Spain, and Sweden register high average values for inclusiveness.

Column (1) of Table 5 highlights the output that resulted from estimating equation (3). As far as the control variables are concerned, the following results are worth taking into account. Economic growth, reflected through gross domestic product per capita, was found to positively influence inclusiveness, although to a modest extent, between 0.07 to 0.12 percentage. Industrial upgrading, represented by the share of value added to the GDP by the services sector, positively determines the degree of inclusiveness within a country, but again to a relatively low extent. An increase by 1 percentage in industrial upgrading leads to an increase in the constructed inclusiveness index of approximatively 0.036 percentages. As such, *H6* has been confirmed.

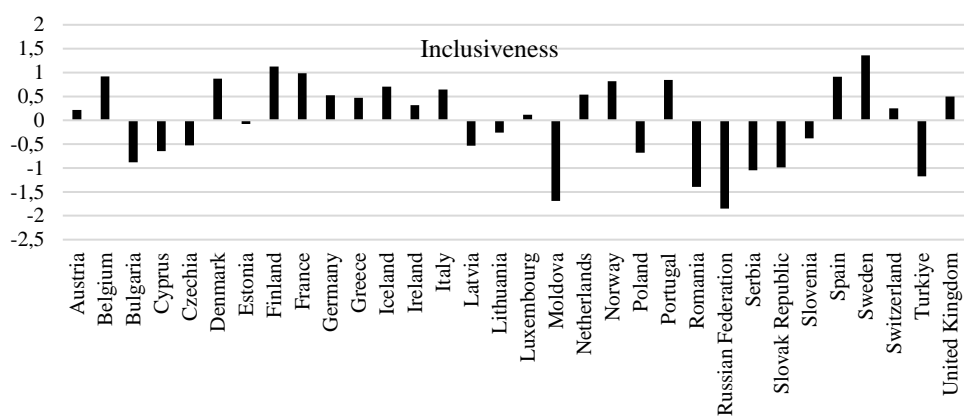

**Figure 1. Average scores for the inclusiveness index of the analysed countries, 2000-2021**
Source: Author's construct based on the existing data, 2023

Furthermore, intersecting with the environmental side of inclusiveness, the quantity of electricity consumed shows a positive influence toward the dependent variable. An increase by 1 percentage for the electricity consumption per capita outlines an increase of roughly 0.046 percentages in the degree of inclusiveness, thus confirming *H1*.

With regard to the social dimension of inclusiveness, the development of the human capital accounts for both positive and negative influences. When taken together with inflation and $CO_2$ emissions (columns 2 and 7) it depicts a positive impact and a negative impact when employed together with digitalisation and governance (columns 1, 3, 4 and 6). To some extent, this finding can be paired with the fact that in the composition of the inclusiveness index, there have been included two indicators that reflect education: population with at least secondary education, male and female. Therefore, given the large number of observations, it is possible that countries with various levels of human capital development may account for similar levels for the two indicators mentioned above. To this end, *H4* has been rejected.

**Table 5. Generalised method of moments (dependent variable: Inclusiveness Index)**

|  | (1) | (2) | (3) | (4) | (5) | (6) | (7) |
|---|---|---|---|---|---|---|---|
| GDP | 0.0711* | 0.1170* | 0.1143* | 0.1181* | 0.1111* | 0.1196* | 0.1063* |
|  | (0.0013) | (0.0012) | (0.0012) | (0.0012) | (0.0012) | (0.0012) | (0.0012) |
| UPGR | 0.0311* | 0.0461* | 0.0432* | 0.0293* | 0.0313* | 0.031* | 0.0416* |
|  | (0.0006) | (0.0005) | (0.0005) | (0.0006) | (0.0006) | (0.0006) | (0.0005) |
| EC | 0.0503* | 0.0415* | 0.0402* | 0.0464* | 0.0450* | 0.0458* | 0.0435* |
|  | (0.0012) | (0.0011) | (0.0011) | (0.0011) | (0.0011) | (0.0011) | (0.0012) |
| HC | -0.0859* | 0.0051* | -0.0029* | -0.0012* | 0.0026* | -0.0037* | 0.0023* |
|  | (0.001) | (0.0004) | (0.0003) | (0.0003) | (0.0003) | (0.0003) | (0.0003) |
| DIG | 0.1121* |  |  |  |  |  |  |
|  | (0.0006) |  |  |  |  |  |  |
| INFL |  | -0.0026* |  |  |  |  |  |
|  |  | (0.0006) |  |  |  |  |  |
| GOV |  |  | -0.0233* |  |  | -0.0126* |  |
|  |  |  | (0.0008) |  |  | (0.0008) |  |
| GCE |  |  |  | 0.0226* |  | 0.0200* |  |
|  |  |  |  | (0.0005) |  | (0.0005) |  |
| GOV*GCE |  |  |  |  | 0.0120* |  |  |
|  |  |  |  |  | (0.0003) |  |  |
| $CO_2$ |  |  |  |  |  |  | -0.0027* |
|  |  |  |  |  |  |  | (0.0004) |
| INCL(-1) | 0.8377* | 0.9097* | 0.9082* | 0.9162* | 0.9115* | 0.9160* | 0.9055* |
|  | (0.0007) | (0.0005) | (0.0005) | (0.0005) | (0.0005) | (0.0005) | (0.0005) |
| C | -0.0102* | -0.1429* | -0.0927* | -0.1624* | -0.1442* | -0.1416* | -0.1153* |
|  | (0.0025) | (0.0022) | (0.0024) | (0.0025) | (0.0025) | (0.0028) | (0.0023) |
| Adjusted R-squared | 0.99 | 0.99 | 0.99 | 0.99 | 0.99 | 0.99 | 0.99 |
| Prob(J-statistic) | 0.08 | 0.00 | 0.00 | 0.00 | 0.00 | 0.00 | 0.00 |
| Durbin-Watson stat | 1.55 | 1.62 | 1.62 | 1.63 | 1.63 | 1.63 | 1.60 |
| Observations | 640 | 640 | 640 | 640 | 640 | 640 | 640 |
| Instruments | 10 | 15 | 15 | 15 | 15 | 15 | 15 |
| Countries | 32 | 32 | 32 | 32 | 32 | 32 | 32 |

Source: Author's construct based on the existing data, 2023

The main findings for the principal determinants selected are as follows. Digitalisation, reflected by the extent of access to internet, proves to be a positive determinant of inclusiveness. This is due to the fact that digitalisation serves as an enhancement of the populations' ability to access information and telecommunication technologies (Ozturk et al., 2022; Wu et al., 2021). Furthermore, digitalisation has proven to generate inclusive growth by mitigating the extent of pollution (Ding et al., 2023; Wang et al., 2022). Based on the current findings, *H5* has been accepted.

Further on, inflation's negative effect on inclusiveness materialises in the fact that an increase of 1 percentage leads to a decrease of 0.0026 of the latter, thus confirming *H7*. Although not a significant impediment to achieving inclusiveness at the European level, the negative effects of inflation may be more pronounced in developing countries (Ozturk et al., 2022).

Contrary to the expected results, governance, reflected through the six qualitative dimensions of the world governance indicators, appears to have a negative, rather than a positive, impact on inclusiveness (columns 3 and 6). An increase of 1 percentage in this indicator appears to lead to a decrease of 0.0233 and 0.0126, respectively, in inclusiveness. This finding highlights the fact that the qualitative dimension of governance is not a positive determinant of inclusiveness, at least not at the European level. On the other hand, the quantitative dimension of governance, reflected through government consumption expenditure materialises its positive influence both when taken separately (column 4) and when taken together with the qualitative dimension (columns 5 and 6). When the product of the two dimensions is employed in the GMM analysis, an increase of 1 percentage in the function of the two indicators leads to an increase of 0.012 percentage in inclusiveness. On the contrary, in column 6, when the two dimensions are taken individually within the same GMM model, they manifest an impact of -0.0126 and +0.0200, respectively. As a consequence, the quantitative dimension of governance proves to be a more pronounced determinant of inclusiveness in developed countries. Therefore, *H3* is rejected, as only the quantitative side of governance exerts an uncontested positive influence on inclusiveness.

The situation for the fourth principal determinant used in determining inclusiveness shows a negative coefficient for $CO_2$ emissions. This suggests that at European level, the extent of pollution's negative effects outpace those that may be considered positive, such as, the expected economic growth generated by the consumption of resources. Therefore, an increase of 1 percentage in the quantity of $CO_2$ emissions leads to a decrease in inclusiveness by 0.0027, thus confirming *H2*.

Thus, developed countries ought to invest in renewable and emission free technologies (Bayar et al., 2019), so that developing countries' future objectives regarding inclusiveness and sustainability are not hampered. Nevertheless, $CO_2$ emissions' impact on the population's welfare are outlined in indicators such as the degree of exposure to PM2.5, the number of deaths caused by such particles and the costs generated in order to mitigate their negative effects (Ofori et al., 2022).

## 5. Conclusions

Motivated by studies that address the determinants of inclusive, green, digital, and sustainable growth in various countries of different levels of economic development, the main goal of the current study was to provide a framework for analysing the determinants of inclusive growth, which can be used by researchers and policymakers to better understand the factors that contribute to inclusive growth. Empirically, the present study is composed of 32 European countries for the period of 2000 to 2021.

In this study, the determinants of inclusiveness and the nature of their influence have been successfully identified. Economic growth, reflected through gross domestic product per capita and by industrial upgrading, was found to positively influence inclusiveness, in line with the findings of Ofori et al. (2022). The quantity of electricity consumed shows a positive influence toward the dependent variable. This finding is consistent with that of Koengkan et al. (2021), who found that the quantity of electricity consumed per capita has a positive influence on the degree of inclusiveness. Digitalisation, reflected by the extent of access to Internet, proves to be a positive determinant of inclusiveness. This finding is consistent with that of Wu et al., (2021), Shodiev et al., (2021) and Kwiliński et al., (2020), who also found that the level of digitalisation has a positive impact on inclusiveness.

The development of the human capital accounts for both positive and negative influences. When taken together with inflation and $CO_2$ emissions, it depicts a positive impact and a negative impact when employed together with digitalisation and governance. This outcome is contrary to that of Wang et al. (2022) or Oyinlola et al. (2021) who found that human capital positively correlates with inclusive growth. Thus, its direct impact on inclusiveness in the case of developed countries needs further research. Contrary to the expectations, the degree of governmental effectiveness reflected through the six dimensions of the World Governance Indicators shows a negative coefficient when employed either alone or paired additively with government consumption expenditure. This highlights the fact that in developed countries, the achievement of inclusiveness standards is to some extent independent of the government's efficiency. On the other hand, when paired multiplicatively with government consumption expenditure (column 5), it depicts a positive coefficient, thus suggesting that the qualitative dimension of governance has to be paired with the quantitative one in order to generate positive results in developed countries. Since only the quantitative side of governance exerts an uncontested positive influence on inclusiveness, this outcome is contrary to that of Ofori et al. (2022), Wang et al. (2022), or Oyinlola et al. (2020), who found that all dimensions of governance have a direct positive effect on inclusive growth.

Given that pollution has effects that occur in both short and long periods, the inclusion of the negative aspects related to the air-polluting particles in the inclusiveness index has attributed a negative coefficient to the principal determinant represented by $CO_2$ emissions. These results reflect those of Koengkan et al. (2021), Asongu, (2018), Asongu et al. (2019), who also found that increasing $CO_2$ emissions have a negative impact on inclusiveness. The findings also show the inflation's negative effect on inclusiveness, in line with the results of Ofori et al. (2022).

The practical implication of this paper is that economic growth, industrial upgrading, the quantity of electricity consumption, and digitalisation are important factors that positively affect inclusive growth in European countries. This data can be used to create targeted interventions that aim to create policies that promote inclusive growth. The limitations of this study arise from the fact that the conducted analyses do not contain the most comprehensive set of indicators. This is because data regarding socio-economic and innovational indicators were not available online for the selected cross-sections. For example, the indicator depicting the income discrepancies could not be included due to missing data. The same is valid for indicators reflecting the innovational context within the selected countries. Since data were not entirely available in online databases, a more comprehensive index of inclusiveness could not be constructed. To a similar extent, data were not available for other indicators that could have been employed in the Generalised Method of Moments analysis as control variables or as principal determinants. Therefore, studies in the field of inclusiveness and sustainability will substantially benefit if indicators such as: education, health, research, and development expenditure as percentage of the gross domestic product will have data update online. Future research could analyse the situation of inclusiveness in the years following the Russo-Ukrainian war to better quantify its impact. The same statement can be applied to the COVID-19 pandemic's aftereffects, as some studies indicate that its long-term negative effects have not yet subsided.